\documentclass{aa}
\usepackage{psfig}
\usepackage{aabib99}
\usepackage{amssymb}

\newcommand\etal{et al.}

\newcommand\lya{Ly$\alpha$}
\newcommand\civl{\hbox{C~\small$\rm IV$}~$\lambda\lambda$~1549}
\newcommand\heiil{\hbox{He~\small$\rm II$}~$\lambda$~1640}
\newcommand\nvl{\hbox{N~\small$\rm V $}~$\lambda$~1240}
\newcommand\siivl{\hbox{Si~\small$\rm IV$}~$\lambda\lambda$~1400}
\newcommand\oiiil{\hbox{O~\small$\rm III]$}~$\lambda$~1665}

\newcommand\civ{\hbox{C~\small$\rm IV$}}
\newcommand\heii{\hbox{He~\small$\rm II$}}
\newcommand\nv{\hbox{N~\small$\rm V $}}
\newcommand\siiv{\hbox{Si~\small$\rm IV$}}
\newcommand\oiii{\hbox{O~\small$\rm III]$}}

\newcommand\aap{{A\&A}}

\newcommand\aj{{AJ}}
\newcommand\apj{{ApJ}}

\newcommand\apjs{{ApJS}}
\newcommand\mnras{{MNRAS}}

\begin{document}


\title{Deep VLT spectroscopy of the $z=2.49$ Radio Galaxy MRC~2104--242\thanks{Based on observations at the ESO VLT Antu telescope}}

\subtitle{Evidence for a metallicity gradient in its extended emission line region}

\author{R.A. Overzier
        \and
        H.J.A. R\"{o}ttgering
        \and
        J.D. Kurk
        \and
        C. De Breuck}

\offprints{R.A. Overzier}

\institute{Sterrewacht Leiden, P.O. Box 9513, 2300 RA, Leiden, The Netherlands (overzier@strw.leidenuniv.nl)}

\date{Received / Accepted}

\maketitle

\begin{abstract}

We present spectroscopic observations of the rest-frame UV 
line emission around radio galaxy MRC 2104--242 
at $z=2.49$, obtained with FORS1 on VLT Antu. 
The morphology of the halo is dominated by two spatially resolved 
regions. \lya\ is extended by $>$12\arcsec\ along the radio axis, 
\civ\ and \heii\ are extended by $\sim$8\arcsec.
The overall spectrum is typical for that of high redshift radio 
galaxies. The most striking spatial variation is that \nv\ is 
present in the spectrum of the region associated with the center 
of the galaxy hosting the radio source, the northern region, while 
absent in the southern region. Assuming that the gas is photoionized 
by a hidden quasar, the difference in \nv\ emission can be explained 
by a metallicity gradient within the halo, with the northern region 
having a metallicity of Z\ $\approx$\ 1.5 Z$_{\odot}$\ and 
Z\ $\leq$\ 0.4 Z$_{\odot}$\ for the southern region. This is consistent with 
a scenario in which the gas is associated with a massive cooling flow or 
originates from the debris of the merging of two or more galaxies.
 
\end{abstract}


\section{Introduction}

High redshift (i.e. $z\gtrsim1$) radio galaxies (HzRGs) are believed to be the 
progenitors of massive elliptical galaxies (e.g. Best et al. 1998).    
\nocite{best98} Therefore, these galaxies 
are an important tool for studying the epoch of galaxy formation 
in the early universe. 
HzRGs are often surrounded by giant 
halos of ionized gas, which radiate luminous emission lines in the UV/optical 
part of the spectrum. The continuum and the line emission, which can be 
spatially extended by as much as 100 kpc, are often elongated along the 
direction of the radio axis (Chambers et al. 1987; for a review see 
McCarthy 1993).\nocite{cham87,mc93} 

One of the most important questions in studying HzRGs 
concerns the ionization of the halo. 
There is striking evidence that the mechanism for ionization is either 
photo-ionization by the active nucleus \cite{vm97} or shock ionization 
by jets interacting with the gaseous medium \cite{best00}. 
Another important question involves the origin of the emission line gas.
If its origin is external to the radio 
galaxy, it can be gas associated with galaxy merging \cite{heck86} or 
the result of a massive cooling flow from the intracluster medium 
\cite{crawfab96}. Alternatively, the gas could have been driven out by a 
starburst-wind or by shocks associated with the radio source. Studying 
the properties of the gas in detail may help to make a distinction between 
these scenarios.  

In this letter we present spectroscopic   
observations of the extended emission line halo around MRC 2104--242.
This radio source is identified with a galaxy at $z=2.49$ 
and is one of the brightest known HzRGs in \lya\ \cite{mc90}. 
Narrowband \lya\ images show a total extent 
of\ $>$ 12\arcsec\ 
(i.e. 136 kpc\footnote[1]{We adopt H$_{0}=50$ 
km s$^{-1}$ Mpc$^{-1}$, q$_{0}=0.1$. At $z=2.49$ this implies a 
linear size scale of 11.3 kpc arcsec$^{-1}$.}) 
distributed in two distinct regions separated by\ 
$\sim$6\arcsec. 
Spectroscopy shows that both regions have large FWHM 
($\sim$$1000-1500$\ km s$^{-1}$), large rest-frame
equivalent widths (330 and 560 \AA) and a velocity difference 
of $\sim$500 km s$^{-1}$ \cite{mc96,koek96,vil99a}. 
The two regions also emit other lines and faint continuum. 
HST imaging (WFPC2 and NICMOS) has shown that the host galaxy 
is very clumpy, suggestive of a merging system (see Pentericci et al. 1999 
for an optical-radio overlay). 
One of the bright components hosted by the northern region is prominent 
in near-infrared emission and therefore it is assumed to be 
the center of the galaxy hosting the radio source. This nucleus and the other 
components in this region are not resolved in our spectra, so we will refer to 
the whole as the northern region.  
The southern region is associated with a narrow filamentary component of 
$\sim$2\arcsec\ oriented in the direction of the radio axis.

The outline of this letter is as follows.
In \S2, we describe our VLT observations and we 
present the basic results in \S3. In \S4, we show evidence 
for a metallicity gradient between the two regions and in \S5 we will 
discuss how this relates to the origin of the halo. 

Throughout this letter we shall abbreviate the emission lines as follows: 
\nv\ for \nvl, \civ\ for \civl, \heii\ for \heiil, \siiv\ for \siivl\ and 
\oiii\ for \oiiil.      


\section{VLT observations}

The observations were carried out in service mode on UT 1999 
September 2$-$5 with FORS1 on the 8.2m VLT Antu 
telescope (ESO-Chile). We used 
grism 600B with a 1\arcsec\ wide slit. A 2$\times$2 readout binning 
was used in order to increase the signal-to-noise ratio (S/N). 
The resultant spectral resolution was $\sim$6 \AA\ (FWHM). 
The slit was positioned along the brightest components and the filamentary 
structure at a position angle of 2$^{\circ}$\ North through East. 
The exposure time was 3$\times$3600s. The seeing during 
the observations was\ $\sim$1\arcsec\ and conditions were photometric. 
Data reduction followed the standard procedures using the NOAO 
IRAF long-slit package. We bias-subtracted the individual frames 
and divided them by a normalized dome flat-field frame. Cosmic rays were 
removed from the background subtracted images. 
We subtracted sky lines and shifted the images 
into registration using stars on the CCD. For 
wavelength calibration we used comparison spectra of a He and a HgCd 
lamp. For flux calibration we observed the spectrophotometric 
standard star LTT7987. The resulting photometric scale is believed to be 
accurate at a level of\ $\sim$15 \%. We corrected the spectra for 
atmospheric extinction and applied a galactic extinction correction 
of E(B$-$V)$=$0.057 determined from the dust maps of Schlegel, Finkbeiner 
\& Davis \cite*{schl98}.

The extraction apertures centered on the two regions were 
resp. 4\arcsec\ and 3\arcsec, chosen to include most of the emission 
while keeping high S/N for the weaker lines. We measured wavelengths, 
fluxes, FWHM and equivalent widths (EW) by fitting Gaussian 
profiles to the lines \cite{rott97}. The measured FWHM was deconvolved for 
the instrumental profile assuming Gaussian distributions.


\section{Results}

The main observational results can be summarized as follows:\\
1. Figs. 1 and 2 show the spectra of the northern and southern regions. 
Both regions show \lya, \civ\ and \heii\ and weak \siiv, \oiii. 
\nv\ is detected in the northern region, but it is absent in the southern. 
The emission line properties are listed in Table 1. 
For the undetected \nv\ in the southern region we have calculated an upper limit
assuming a Gaussian line shape with a peak 3 times the rms of the
continuum at the expected wavelength and a width comparable to that of
\nv\ in the northern region.\\
2. Fig. 3 shows the two-dimensional emission line structure of \lya, \civ\ 
and \heii. The peak of the \civ\ emission in the northern region is redshifted 
with respect to that of \lya\ by $\sim$100 km s$^{-1}$, while that of \heii\ 
is blueshifted by $\sim$150 km s$^{-1}$.
In the southern region, \lya\ shows two separate peaks shifted blueward from 
the northern region by $\sim$1000 and $\sim$500 km s$^{-1}$. This two-peak 
distribution is also seen in \civ\ and \heii, albeit at low S/N. The fact that it 
is observed in \heii\ could indicate kinematical substructure 
in the halo and that the dip in the \lya\ profile is not due to H I absorption.\\   
3. The northern region shows similar FWHM ($\sim$700 km s$^{-1}$) for \lya, \civ\ and \heii. 
In the southern, \lya\ and \civ\ have high FWHM ($>$1000 km s$^{-1}$), 
while that of \heii\ is a factor 1.5 lower. \\    
4. Within the errors, the emission line ratios of the two regions are the same, 
only those involving \nv\ are discrepant. The \nv/\civ\ and \nv/\heii\ line ratios 
are at least 4 and 3 times higher in the northern region compared to the 
southern. 

\section{Difference in \nv\ emission from the two regions: evidence for a 
metallicity gradient?}

We have detected \nv\ in the northern region, which is seldom present at a 
significant level in HzRGs \cite{rott97,db00}. In the southern region we 
detected no \nv, while the other line-ratios are similar in both regions. 
Vernet et al. (1999) found that HzRGs follow a sequence in \nv/\civ\ vs. 
\nv/\heii, parallel to the relation defined by the broad line 
regions (BLR) of quasars found by Hamann \& Ferland (1993).\nocite{hamfer93} 
These authors showed that this sequence can be explained by a variation of the 
metallicity of the BLR, caused by a rapidly evolving starburst in the 
massive galactic core. Villar-Mart\'{\i}n et al. (1999) \nocite{vil99b} 
showed that both shock ionization and photo-ionization could not explain 
either the \nv\ correlation, or the strong \nv\ emission in some HzRGs 
(e.g. van Ojik \etal\ 1994).\nocite{ojik94} They found that a model of 
photo-ionization and variation of metallicity best explained both properties. 
Therefore, we conclude that the difference in \nv\ emission can be 
explained only by a metallicity gradient within the halo. Using the 
metallicity sequence with quadratic nitrogen enhancement 
(N\ $\propto$\ Z$^{2}$) from Vernet et al. \cite{ver00} we find a 
metallicity of Z\ $\approx$\ 1.5 Z$_{\odot}$ for the 
northern region and an upper limit of Z\ $\leq$\ 0.4 Z$_{\odot}$\ for the 
southern (Fig. 4).

\section{Discussion}

The supersolar metallicity of the gas associated with the central 
part of the galaxy implies that 2104--242 has experienced a period of 
intense star formation. Assuming that 
the difference in \nv\ emission 
found within the halo is due to a metallicity gradient, it is likely 
that the emitting 
\newpage

\begin{figure*}[t]
\begin{minipage}[t]{8.8cm}
\psfig{figure=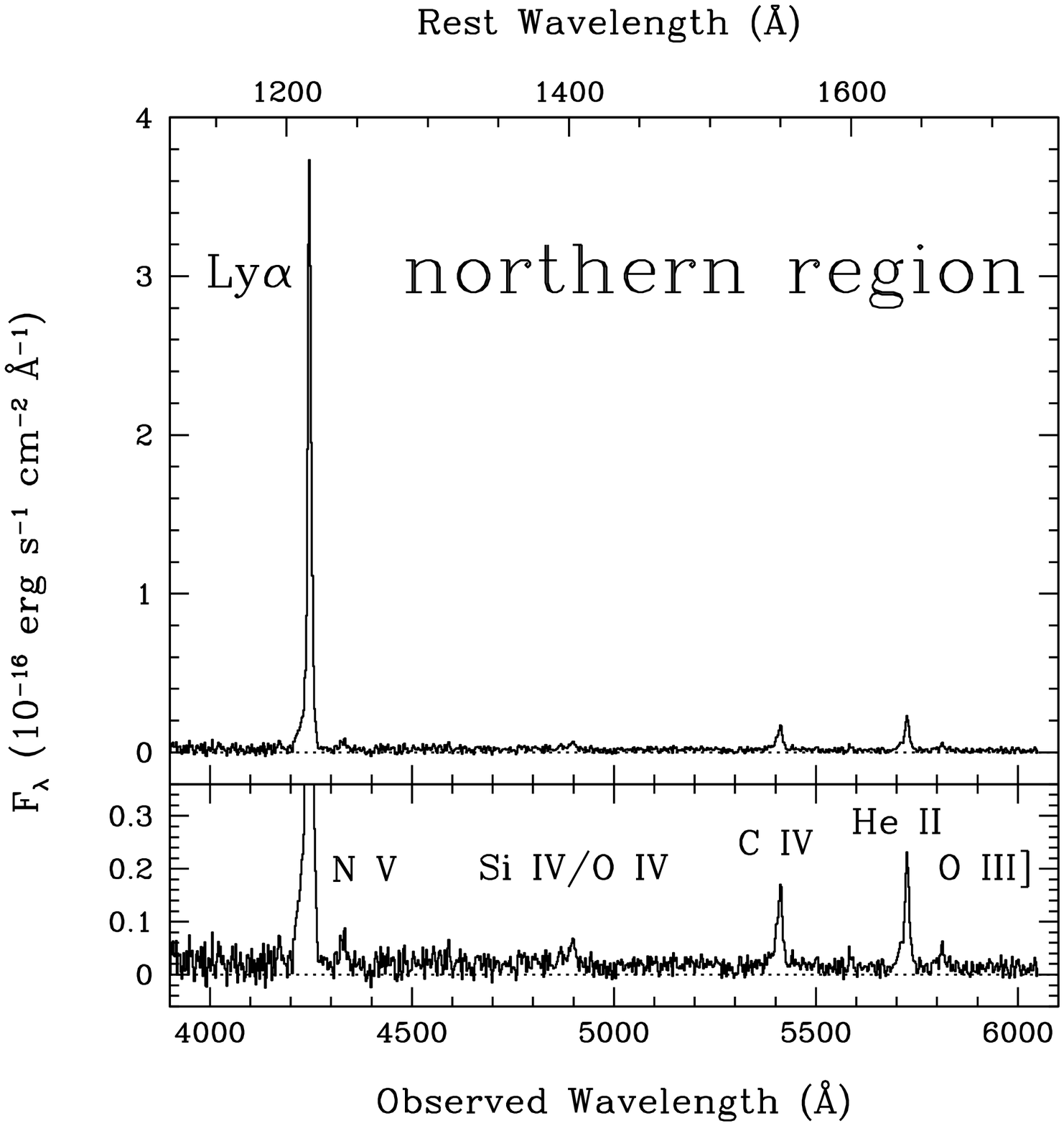,width=8.8cm}
\caption[]{VLT/Antu spectrum of the northern region of emission 
of 2104--242. The extraction aperture was 4\arcsec.}
\end{minipage}
\hfill
\begin{minipage}[t]{8.8cm}
\psfig{figure=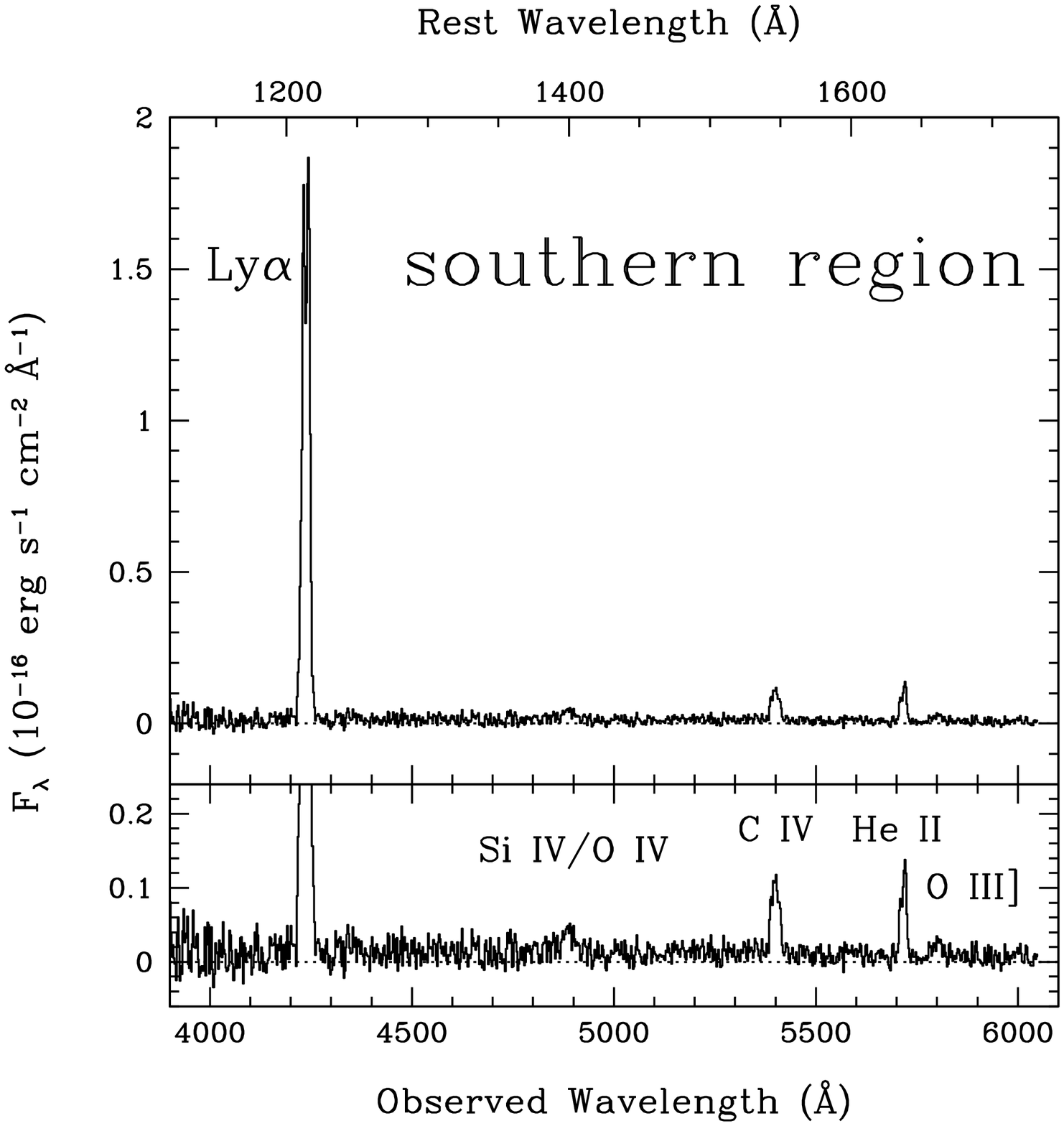,width=8.8cm}
\caption[]{VLT/Antu spectrum of the southern region of emission 
of 2104--242. The extraction aperture was 3\arcsec.}
\end{minipage}
\end{figure*}

\begin{figure*}[ht]
\begin{minipage}[t]{18cm}
\hbox{\hspace{0cm}
\psfig{figure=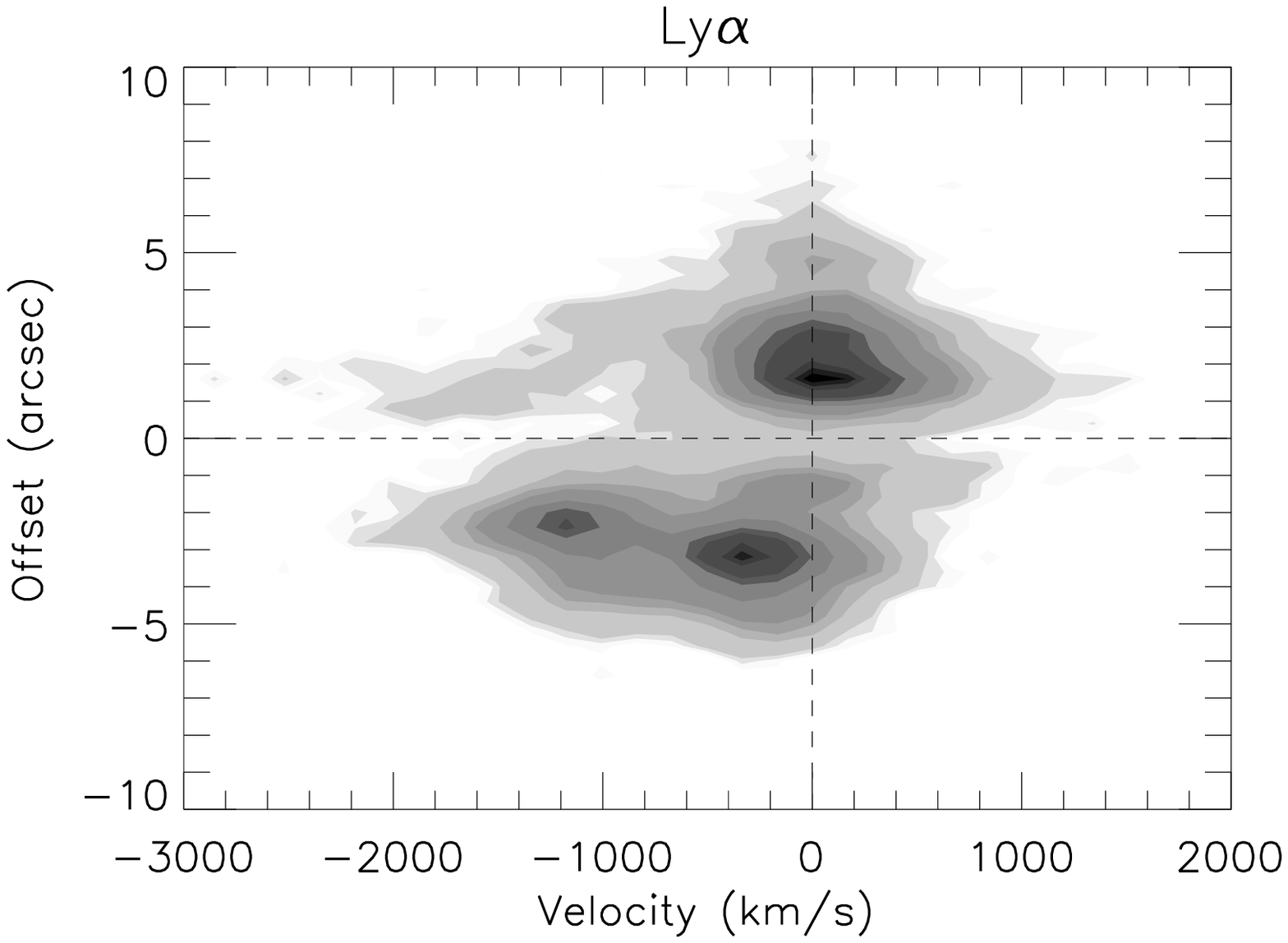,width=6cm,height=7cm}\hspace{0cm}
\psfig{figure=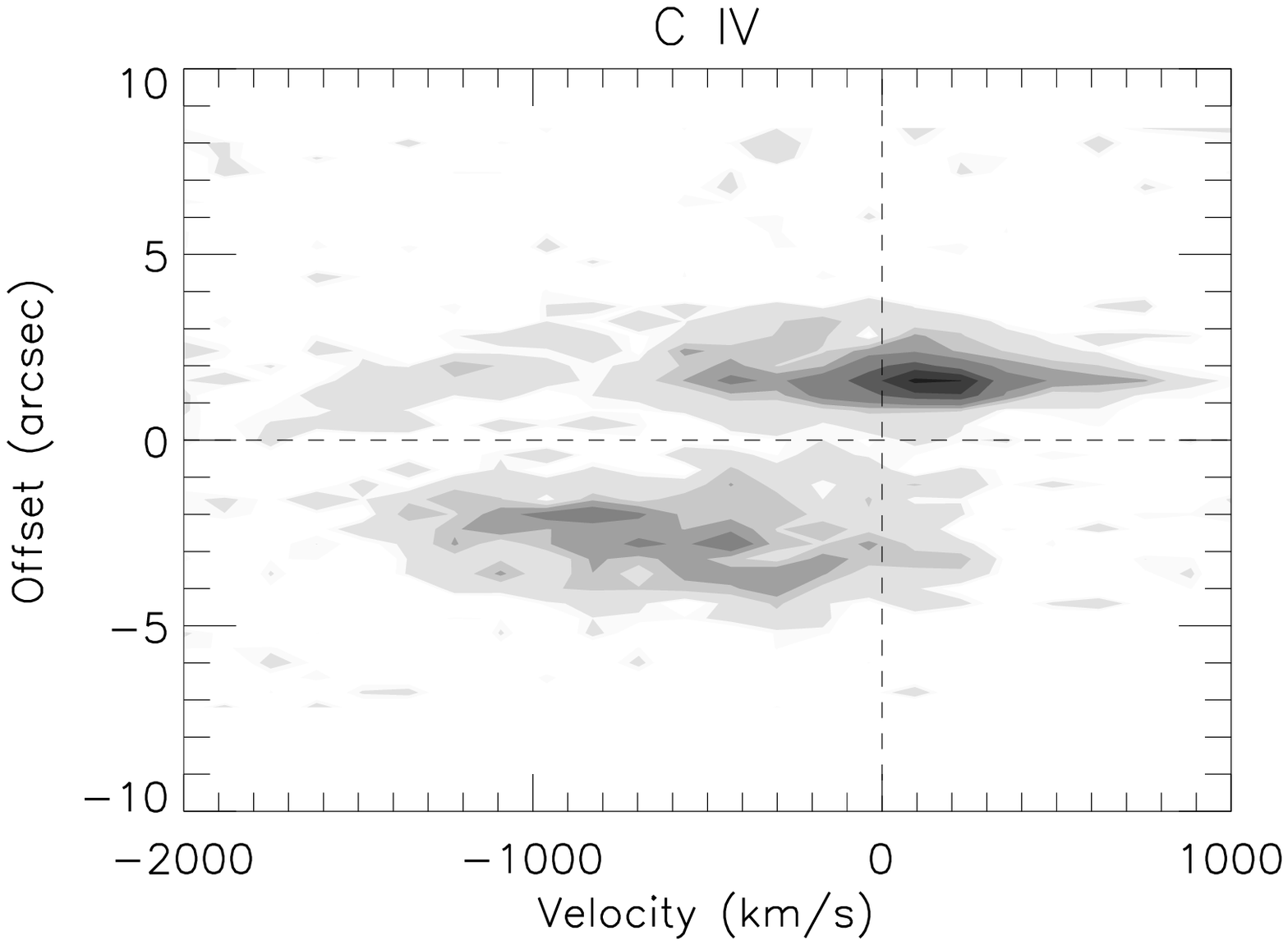,width=6cm,height=7cm}\hspace{0cm}
\psfig{figure=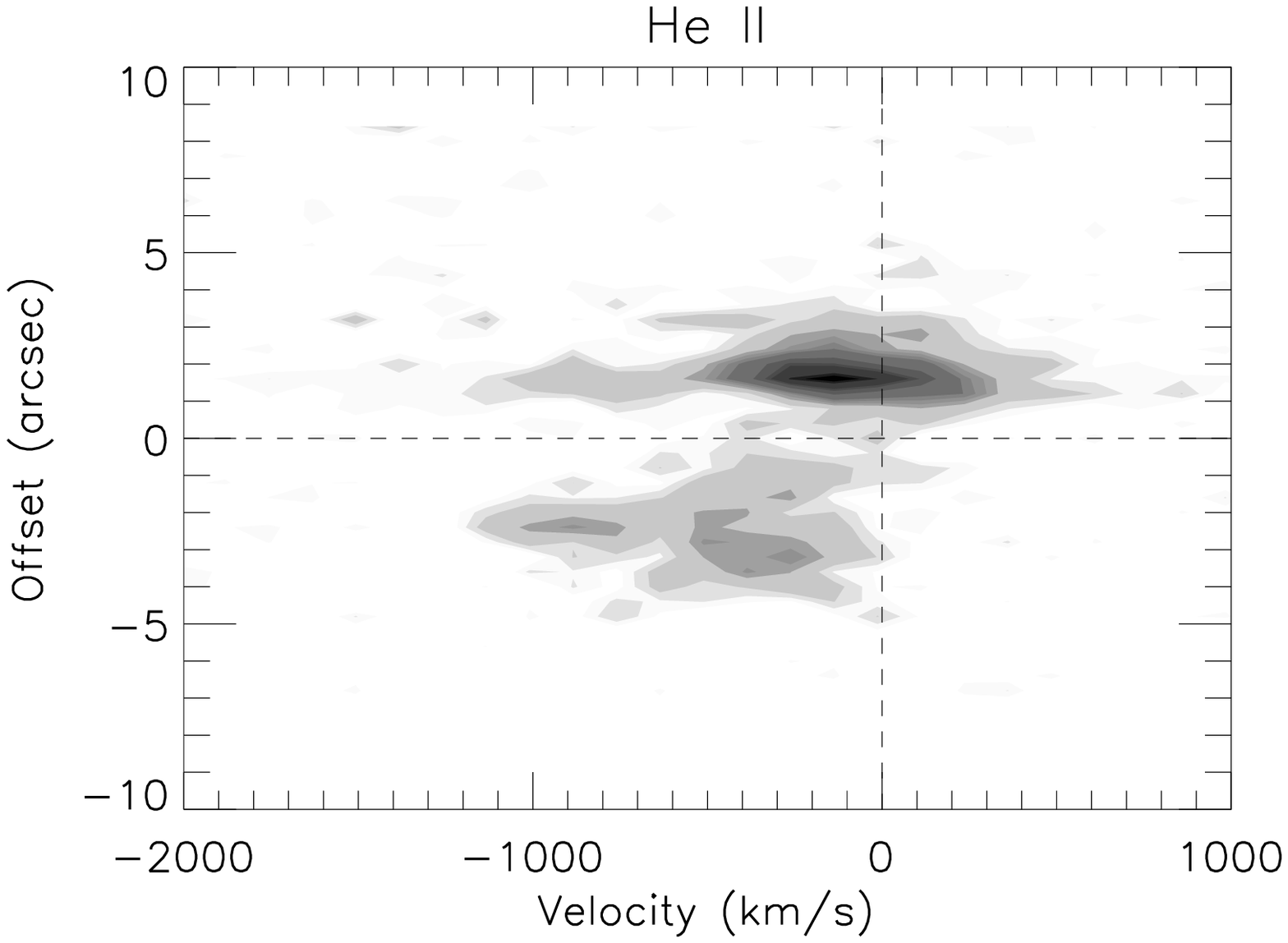,width=6cm,height=7cm}}
\caption{The two-dimensional emission line structures of \lya, \civ\ and \heii. Offset zero was chosen in between the northern and southern regions. Velocity zero corresponds to the peak \lya\ emission in the northern region.}
\hfill
\end{minipage}
\end{figure*}

\begin{table*}[ht]
\begin{minipage}[t]{18cm}
\caption[]{Wavelength, flux, FWHM and (rest-frame) EW for the emission lines in the northern and southern regions.}
\begin{tabular}{l|llll|llll}
\hline
\multicolumn{1}{}{} &
\multicolumn{4}{c|}{northern region} & 
\multicolumn{4}{|c}{southern region}\\
\hline

Line       & Peak (\AA) & Flux$^{a}$  &FWHM (km s$^{-1}$) & EW (\AA) & Peak (\AA) & Flux$^{a}$ & FWHM (km s$^{-1}$) & EW (\AA)\\
\hline
\lya       & $4247\pm1$ & $40\pm4$    & $610\pm140$   & $360\pm50$  & $4238\pm1$ & $42\pm4$    & $1490\pm140$ & $845\pm232$\\
\nv        & $4330\pm2$ & $1.0\pm0.1$ & $1100\pm300$  & $13\pm2$    &            & $\le0.25$   &              & \\
\siiv & $4892\pm8$ & $1.2\pm0.1$ & $2200\pm1000$ & $18\pm4$    & $4887\pm5$ & $1.2\pm0.2$ & $2300\pm700$ & $32\pm5$\\
\civ       & $5411\pm1$ & $2.4\pm0.3$ & $850\pm150$   & $42\pm6$    & $5399\pm2$ & $2.5\pm0.3$ & $1200\pm300$ & $57\pm17$\\
\heii      & $5726\pm1$ & $3.0\pm0.3$ & $700\pm100$   & $55\pm7$    & $5717\pm2$ & $2.0\pm0.2$ & $800\pm200$  & $58\pm22$\\
\oiii     & $5812\pm1$ & $0.3\pm0.1$ & $250\pm200$   & $4\pm1$     & $5803\pm7$ & $0.3\pm0.1$ & $750\pm700$  & $6\pm2$\\

\hline
\multicolumn{8}{l}{$^a$ Flux is given in units of 10$^{-16}$ erg s$^{-1}$ cm$^{-2}$ }\\
\end{tabular}
\end{minipage}
\end{table*}

\clearpage
\newpage


\begin{figure}[t]
\psfig{figure=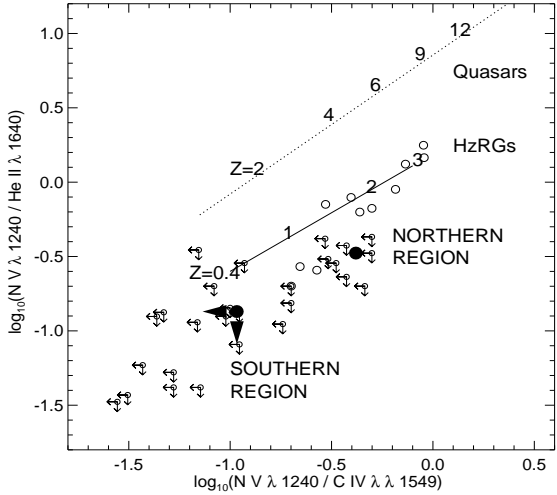,width=8cm,height=7cm}
\caption[]{\nv/\heii\ vs. \nv/\civ. The dotted line represents the 
metallicity sequence defined by quasars (Hamann \& Ferland 1993), 
with the numbers along the line indicating the metallicity in solar units. 
The solid line represents a metallicity sequence with N\ $\propto$\ Z$^{2}$ 
(ionization parameter U$=$0.035, power law spectral index $\alpha$$=-$1.0) 
from Vernet et al. (2000). The two regions of 2104--242 are 
indicated. Small open circles indicate radio galaxies from the sample of 
De Breuck et al. (2000).}
\end{figure}

\noindent
gas near the center and the gas further out are in different
stages of evolution.

The infall of gas by massive cooling flows is believed to 
be an important process in galaxy formation (Crawford \& Fabian 1996).
\nocite{crawfab96} In this scenario, gas cools from a primordial 
halo surrounding the radio source and provides the material from which the 
galaxy is made. This could also be the case for 2104--242. 
Fardal et al. (2000) \nocite{fardal00} recently examined 
cooling radiation from forming galaxies, focusing on \lya\ line 
luminosities of high redshift systems. They find that a significant amount 
of the extended \lya\ emission can arise from cooling radiation. 

However, because the two regions of 2104--242 both show 
emission lines from other elements than H and He, at least some of 
the gas must already have been processed in stars in the past. 
The HST images revealed that this galaxy actually 
consists of a number of larger and smaller components \cite{pen99}, 
which supports the general idea that galaxies are formed by a process 
of hierarchical buildup \cite{kauf95}. We have also found evidence for 
kinematical substructure within the halo. Therefore, the emission line halo 
may be the result of gas associated with such intensive galaxy merging. 

Alternatively, the gas in the halo may be the result of 
other mechanisms. It may have been driven out by strong jet-cloud 
interactions. Line widths as large as 1500 km s$^{-1}$ are common 
in HzRGs, which indicate extreme, non-gravitational motions. The 
alignment effect also suggests that jet-cloud interactions occur and 
for some sources there is evidence of shock ionization (Best et al. 2000).  
\nocite{best00} However, large radio sources like 2104--242 are less 
likely to be 'shock-dominated', because the shockfronts have passed 
well beyond the emission line halo.  

Also, the gas could have been expelled by a superwind following an 
enormous starburst. We have shown evidence for intense star formation 
in 2104--242. Binette et al. (2000) \nocite{binette00} showed that 
radio galaxy 0943--242\ is surrounded by a vestige gas shell of very low 
metallicity. They conclude that this gas has been expelled from 
the parent galaxy during the initial starburst at the onset of its formation.
 
We conclude that the emission line ratios are well explained by a 
combination of photo-ionization and a metallicity gradient. This is 
consistent with scenarios in which the halo is formed by gas falling 
onto the radio galaxy located at the center of a forming cluster or 
by gas associated with intense galaxy merging. 
   
\begin{acknowledgements}
We acknowledge productive discussions with Wil van Breugel and Laura Pentericci.
\end{acknowledgements}


\begin{thebibliography}{}

\bibitem[Best \etal\ 1998]{best98} Best P., Longair M., R\"ottgering H., 1998, \mnras\ 295, 549

\bibitem[Best \etal\ 2000]{best00} Best P., R\"ottgering H., Longair M., 2000, \mnras\ 311, 23

\bibitem[]{binette00} Binette L., Kurk J., Villar-Mart\'{\i}n M., R\"{o}ttgering H., 2000, \aap\ 356, 23

\bibitem[Chambers et al 1987]{cham87} Chambers K., Miley G., van Breugel W., 1987, Nat 329, 604

\bibitem[De Breuck et al. 2000]{db00} De Breuck C., R{\"o}ttgering H., Miley G., van Breugel W., Best P., 2000, \aap\ 362, 519

\bibitem[]{fardal00} Fardal M. et al., 2000, in press, astro-ph/0007205

\bibitem[Crawford \& Fabian 1996]{crawfab96} Crawford C., Fabian A., 1996, \mnras\ 282, 1483

\bibitem[Hamann \& Ferland 1993]{hamfer93}Hamann F., Ferland G., 1993, \apj\ 418, 11

\bibitem[Heckman \etal\ 1986]{heck86} Heckman T. \etal, 1986, \apj\ 311, 526

\bibitem[Kauffmann 1995]{kauf95}Kauffmann G., 1995, \mnras\ 274, 161

\bibitem[Koekemoer \etal\ 1996]{koek96}Koekemoer A., van Breugel W. \& Bland-Hawthorn J., 1996, In: M. Bremer \etal\ (eds.) ``Cold Gas at High Redshifts'', Kluwer, p. 385

\bibitem[McCarthy \etal\ 1990]{mc90} McCarthy P., Kapahi V., van Breugel W., Subrahmanya C., 1990, \aj\ 100, 1014

\bibitem[McCarthy \etal\ 1993]{mc93} McCarthy P., 1993, ARA\&A\ 31, 639

\bibitem[McCarthy \etal\ 1996]{mc96} McCarthy P., Baum S., Spinrad H., 1996, \apjs\ 106, 281

\bibitem[Pentericci \etal\ 1999]{pen99} Pentericci L. \etal, 1999, \aap\ 341, 329 

\bibitem[R\"{o}ttgering \etal\ 1997]{rott97} R\"{o}ttgering H. \etal, 1997, \aap\ 326, 505 

\bibitem[1998]{schl98} Schlegel D., Finkbeiner D., Davis M., 1998, \apj\ 500, 525

\bibitem[van Ojik \etal\ 1994]{ojik94} van Ojik R. \etal, 1994, \aap\ 289, 54

\bibitem[2000]{ver00} Vernet J. \etal, 2000, in press, astro-ph/0010640

\bibitem[Villar-Mart\'{\i}n \etal\ 1997]{vm97} Villar-Mart\'{\i}n M., Tadhunter C., Clark N., 1997, \aap\ 323, 21

\bibitem[Villar-Mart\'{\i}n \etal\ 1999a]{vil99a} Villar-Mart\'{\i}n M., Binette L., Fosbury R., 1999, \aap\ 346, 7

\bibitem[Villar-Mart\'{\i}n \etal\ 1999b]{vil99b} Villar-Mart\'{\i}n M., Fosbury R., Binette L., Tadhunter C., Rocca-Volmerange B., 1999, \aap\ 351, 47

\end{thebibliography}
\end{document}